\documentstyle[twoside,fleqn,espcrc2,epsf]{article}

\newcommand{\AmS}{{\protect\the\textfont2
  A\kern-.1667em\lower.5ex\hbox{M}\kern-.125emS}}

\title{Constraints on color dipole-nucleon cross section from 
diffractive heavy quarkonium production}

\author{Katsuhiko Suzuki\address[MCSD]{Department of Physics, University of 
Tokyo, Tokyo 113-0033, Japan}%
        \thanks{e-mail: ksuzuki@nt.phys.s.u-tokyo.ac.jp, supported by JSPS}
        and
        Kazunori Itakura\address{RIKEN-BNL Research Center, Brookhaven 
        National Laboratory, Upton, NY 11973, USA}
        \thanks{e-mail: itakura@bnl.gov}
        }
       
\begin{document}

\begin{abstract}
We study the hard color dipole-nucleon cross section within perturbative QCD  
and discuss its relation to observables in diffractive leptoproduction of 
heavy quarkonium.  
The dipole cross section calculated with the unintegrated gluon density of the 
nucleon substantially differs from the well-known perturbative form 
$\sigma_{dip} \sim b^2$ 
for $b > 0.3$fm, where $b$ is the transverse separation of the dipole.  
We show the measured ratio 
of $\psi '$ to $J / \psi$ photoproduction cross sections constrains the dipole 
cross section at intermediate $b$, and in fact excludes the simple 
$\sigma_{dip} \sim b^2$ behavior.    
We also calculate the $t$-slopes of the diffractive $J / \psi, \psi '$ 
productions.  
We emphasize the difference of $t$-slopes, $B_{J/\psi} - B_{\psi '}$, 
is dominated by the dipole-nucleon dynamics.  This difference is found to be
about $0.3 \mbox{GeV}^{-2}$ with our dipole cross section.
\vspace{1pc}
\end{abstract}

\maketitle

\section{Introduction}

Recently hard diffractive processes in QCD have attracted 
considerable theoretical and experimental interests.  
In particular, diffractive photo- and leptoproductions of vector 
mesons off the proton, $\gamma^{(*)} + p \to V(\rho^0, \omega, 
\phi,J/\psi, \dots) + p$, provide crucial constraints on 
gluon dynamics at small-$x$\cite{Brodsky,Ryskin,FKS,Martin,Nemchik}.  
In the target rest frame, this process can be described by the 
color dipole picture, where the incoming photon undergoes 
$q \bar q$ fluctuations and these components subsequently 
interact with the target nucleon\cite{Text}.    
Within this picture, a number of exclusive and inclusive processes 
can be formulated in terms of a {\em universal} color dipole-nucleon 
cross section by virtue of the QCD factorization theorem.

Behavior of the dipole-nucleon cross section $\sigma_{dip} (b)$ is 
of great interest, where $b$ is a transverse size of the dipole, 
especially for intermediate  region~($b\sim $ 0.3-0.4fm).~There~are 
several phenomenological works to determine the 
dipole cross section to reproduce the experimental data\cite{Dipole}.  
In this work, 
we calculate the dipole-nucleon cross section within perturbative 
QCD.   Hence, we deal with the process which contains a hard scale 
such as a large momentum transfer $Q^2,t$ or the heavy quark mass.  
Since we are interesetd in the vector meson production at HERA energy, 
typically $x \sim 10^{-3}$, we do not cosider the saturation of the 
gluon density at small-$x$.\footnote{If we focus on much smaller-$x$, 
e.g.~$x\sim 
10^{-5}$, we should include the saturation effects on the dipole-nucleon 
cross section.}     
We discuss how resulting dipole cross section differs from the 
well-known perturbative $\sigma_{dip} \sim b^2$ behavior, and 
calculate observables which are sensitive to the 
behavior of the dipole cross section.

\section{Hard color dipole cross section}
Leading order diagrams for dipole-nucleon scattering, 
$q \bar q (q) + p (p) \to q \bar q (q+ \Delta) + p (p - \Delta)$,   
are shown in Fig.1. 
With the unintegrated gluon density of the proton $f (x,l_T)$, 
the dipole cross section at forward limit, $t=\Delta^2 =0$, is expressed as
\begin{eqnarray}
\sigma _{dip}={{4\pi ^2\alpha _s} \over 3}\,\int^{Q^2}  {dl_T^2}
\, \left[1-J_0(b\,l_T)\, \right] \frac{f(x,l_T^2)}{l_T^4}
\label{full}
\end{eqnarray}
where $Q^2$ is the virtuality of the incoming photon which sets the 
scale of the process, and $x, l_T$ 
are longitudinal momentum fraction and transverse momentum of the gluon, 
respectively.    
Our calculation is essentially based on the $k_T$-factorization scheme.  
Relation between $Q^2$ and the transverse distance $b$ is studied in 
ref.~\cite{MFGS}.  

\vspace{-0.5cm}

\begin{figure}[htb]
\epsfxsize = 6 cm   
\centerline{\epsfbox{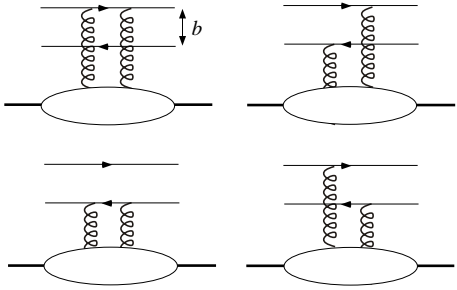}}
\vspace{-0.8cm}
\caption{Diagrams for dipole-nucleon scattering}
\label{fig:fig1}
\end{figure}

\vspace{-0.5cm}

The unintegrated gluon density is not precisely known at present.   
In principle, there is a $Q^2$ dependence in $f(x,l_T^2,Q^2)$, and   
such a scale dependence can be studied, say, 
by solving CCFM equation\cite{unint}, although we do not incorporate 
it in this work.   
Here, we simply assume that the unintegrated gluon density is 
related to the conventional gluon 
distribution measured in deep inelastic scattering via
\begin{eqnarray}
f(x,l_T)
\equiv {l_T^2}  \frac{ \partial \, x G(x,l_T^2)}{\partial \, l_T^2} \; .
\end{eqnarray}

Before going to evaluate eq.~(\ref{full}) 
explicitly, we shall show an approximate 
expression for the small size dipole.  Assuming $b l_T << 1$ and keeping 
terms up to second order of $(b \cdot l_T)$, we rewrite eq.~(\ref{full}) as
\begin{eqnarray}
\sigma _{dip}={{\pi ^2\alpha _s} \over 3}\,
\, b^2 \, xG(x,Q^2)
\label{dipole}
\end{eqnarray}
We refer this approximation as the small dipole approximation (SDA), 
in which the dipole cross section has a simple geometrical expression 
$\sigma_{dip} \sim b^2$.   
This expression is frequently used for various applications\cite{Text}.  
It is clear that this approximation works well for small-$b$ region, or, 
in other words, the process with large $Q^2$.

Now we come back to eq.~(\ref{full}).  
To integrate out eq.~(\ref{full}), we use the
parametrization of the gluon distribution function.   
However, available parametrizations are known for $Q^2 \geq$ a few  
$\mbox{GeV}^2$.  
Here, we follow one of the prescriptions made in ref.~\cite{Levin}:   
the linear extrapolation for $Q^2$ below $Q^2_0$, 
\begin{eqnarray}
G(x,Q^2 ) = \left\{ {\begin{array}{*{20}c}
   {G(x,Q^2 )\quad Q^2  \ge Q_0^2 }  \\
   {\frac{{Q^2 }}{{Q_0^2 }}G(x,Q_0^2 )\quad Q^2  \le Q_0^2 }  \\
\end{array}} \right.
\end{eqnarray}
where $Q^2_0$ is an initial input scale of the parametrizations
$\sim 1 \mbox{GeV}^2$.

Calculated dipole cross section is shown in Fig.2 at $x=10^{-3}$.  
Solid curve denotes our full result eq.~(\ref{full}), while the dashed one 
indicates the result with SDA (\ref{dipole}).   
Both results agree at small $b < 0.2$fm as expected.  
However, difference between them is substantial at $b=0.3 \sim 0.4$fm.  
Since we use the perturbative technique, the result at $b \sim 1$fm 
cannot be justified.  Nevertheless, there is a distinct deviation 
at intermediate $b$, 
which is crucial for observables discussed in the following sections.  
To get the full result of Fig.2, we used CTEQ5m\cite{CTEQ} with the 
linear extrapolation.  We have also used several sets of parametrization of 
the gluon distribution and other approximation scheme for the low $Q^2$ 
region.  
The difference is about $10 \sim 20 \%$.  

\vspace{-0.4cm}

\begin{figure}[htb]
\epsfxsize = 6cm   
\centerline{\epsfbox{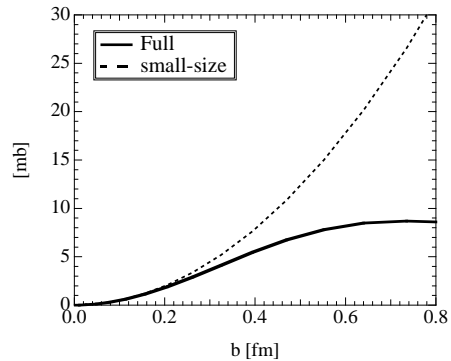}}
\vspace{-0.8cm}
\caption{Color dipole-nucleon cross section}
\label{fig:fig2}
\end{figure}

\vspace{-0.4cm}

\section{$\psi '$ to $J /\psi$ ratio}

Using the dipole cross section, one can write the diffractive 
vector meson production amplitude symbolically as;
\begin{eqnarray}
{\cal A} = \psi_\gamma \otimes \sigma_{dip} \otimes \psi_{V}
\label{amp}
\end{eqnarray}
where $\psi_\gamma, \psi_V$ are light-cone wave functions of the photon and 
vector meson, respectively.   In eq.~(\ref{amp}) integrations are 
performed over $z$, longitudinal momentum fraction of a heavy quark, and $b$.  
Typical scale of this process is $Q^2_{eff} = Q^2 / 4 + m^2$ where 
$m$ is the heavy quark mass.

The SDA approach is adopted to evaluate the $J / \psi$ production 
in refs.~\cite{Brodsky,FKS}, 
and gives reasonable results in agreement with the data.   
However, as pointed out by Hoyer and  Peign{\'e}\cite{Hoyer} very recently, 
the ratio of $\psi '$ to $J / \psi$ photoproduction cross sections 
calculated in SDA is much smaller 
than the measured values\cite{ExpHERA}, when one 
uses realistic non-relativistic quark models to obtain the charmonium wave 
functions.  
The formula used in SDA is obtained by assuming that 
$(b \cdot l_T)$ is small as discussed in section 2.    
However, the quark model indicates that 
the size of $\psi '\sim 0.8 \mbox{fm}$ is twice as large as 
the radius of $J / \psi$.  
Hence, the validity of SDA is questionable 
for $\psi ' (2S)$  case.  
We may  expect considerable contributions from the overlap of the 
large size color dipole and $\psi ' $ wave function.  
Above argument motivates us to calculate the diffractive $J / \psi $ and 
$\psi '$ productions without resorting to the small dipole 
approximation\cite{Suzuki}.

\begin{figure}[htb]
\epsfxsize = 6 cm   
\centerline{\epsfbox{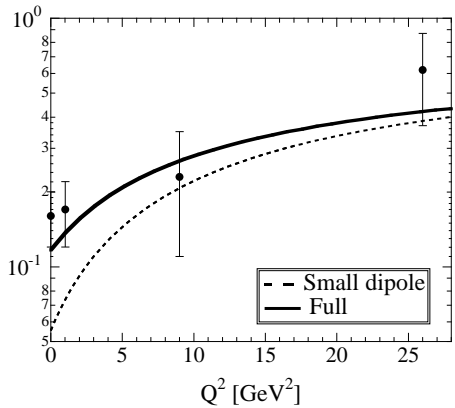}}
\vspace{-0.8cm}
\caption{$\psi '$ to $J / \psi$ production ratio}
\label{fig:fig2}
\end{figure}

We obtain the results shown in Fig.3 with the dipole cross section of 
Fig.2\cite{Suzuki}.  
Calculation without (with) the small dipole approximation is depicted 
by the solid (dashed) curve.   For large $Q^2$ 
the difference between them becomes small as expected.  
However, around $Q^2 = 0$, the full calculation  
with eqs.~(\ref{full}) and (2)  is clearly larger than 
the result obtained by eq.~(\ref{dipole}) by factor about 2, 
and shows a reasonable agreement with the experimental data\cite{ExpHERA}.

We also consider the production of $\Upsilon$-states.   
Results for $\Upsilon ' (2S) / \Upsilon (1S)$ 
and  $\Upsilon '(3S) / \Upsilon (1S)$ are consistent 
with those obtained by SDA.  
SDA is already enough for the 
$\Upsilon$ production due to the large bottom 
quark mass.

\section{$t$-slope of heavy quarkonium production}

Measured diffractive production cross section of the vector meson at 
small momentum transfer, $t\neq 0$, can be parametrized as,
\begin{eqnarray} 
\frac{d \sigma} {dt} = 
\left( \frac{d \sigma} {dt} \right)_{t=0} \cdot \mbox{exp}( B t)
\label{tslope1}
\end{eqnarray}
where the diffractive slope $B$ is a function of $W,Q^2,t$ in general.  
At low $t$, we may decompose the $t$-slope $B$ as 
\begin{eqnarray}
B = B_{dip} + B_{Soft} + \cdots \; , 
\label{tslope2}
\end{eqnarray}
where  $B_{dip}$ indicates contributions from the 
color dipole-nucleon  dynamics, and the remainder expresses other 
possible contributions including the soft nucleon structure $B_{Soft}$.

The contribution from the dipole part for small $t$ is given by\cite{FKS}
\begin{eqnarray}
B_{dip} \propto \psi_\gamma \otimes b^2 \sigma_{dip} \otimes \psi_V \; . 
\end{eqnarray}
Our results for $B_{dip}$ are shown in Fig.4 for $J/ \psi , \psi'$, although 
$B_{dip}$ cannot be directly compared with the data.  Here, $L$ and $T$ 
denote the longitudinal and transverse polarizations of the photon.  
For $J/ \psi$, results with the full dipole cross section and SDA 
give almost the same behavior.  However, for $\psi '$, there is a 
clear difference.  Such large differences come from the 
intermediate-$b$ behavior of the dipole cross section.  In turn, the $t$-slope 
of the $\psi '$ production could be useful to constrain the 
 behavior of the dipole-nucleon cross section for $b > 0.3$fm.

\begin{figure}[htb]
\epsfxsize = 6.8 cm   
\centerline{\epsfbox{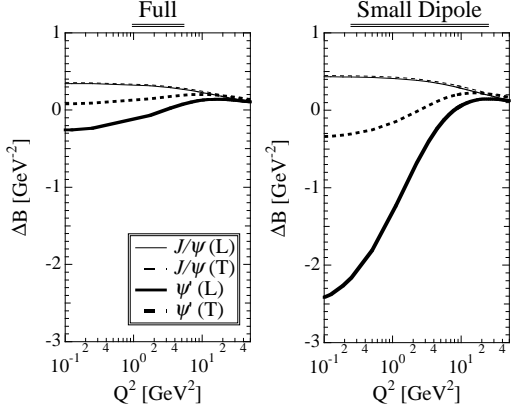}}
\vspace{-0.8cm}
\caption{$B_{nS}^{dip}$ for various photon polarizations}
\label{fig:fig4}
\end{figure}

Let us consider the difference of $t$-slopes, $B_{J/ \psi} - B_{\psi '}$.  
Because the contribution from soft nucleon structure 
in eq.~(\ref{tslope2}) is 
independent of wave functions of the vector mesons, it must be 
the same for $J / \psi$ and $\psi '$ productions.  
Therefore, it is easy to arrive at a relation;
\begin{eqnarray}
B_{J/ \psi } - B_{\psi '} = B_{dip} (J /\psi ) - B_{dip} (\psi ')
\end{eqnarray}
which suggests $B_{J/ \psi } - B_{\psi '}$ is dominated by the dipole 
contribution, and thus a suitable quantity to study the shape of 
the dipole-nucleon cross section.  
We calculate this difference for the transversely polarized case 
shown in Fig.5.  
Results with the full dipole cross section gives a considerably 
smaller value than that of SDA.  Recent preliminary data from 
HERA\cite{psitslope} is rather consistent with our full calculation.  
More accurate data is quite important for detailed studies.

\begin{figure}[htb]
\epsfxsize = 6 cm   
\centerline{\epsfbox{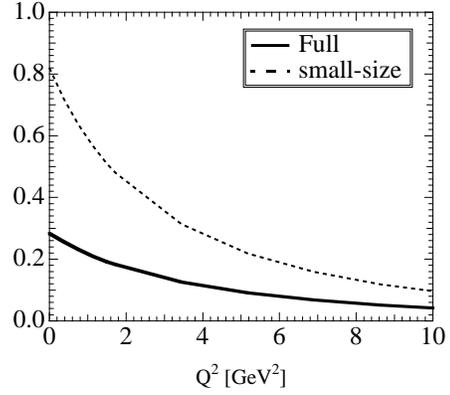}}
\vspace{-0.8cm}
\caption{$B_{J / \psi } - B_{\psi '}$ [GeV$^{-2}$]}
\label{fig:fig5}
\end{figure}

\section{Conclusion}

We have calculated the $q \bar q$ dipole-nucleon cross section in terms of the 
unintegrated gluon density.  The dipole cross section 
is the universal quantity which characterizes the various hard processes.   
We have found the $\psi '$ to $J /\psi$ diffractive production cross 
section ratio
is sensitive to the behavior of $\sigma_{dip}$ for $b>0.3$fm.  
We have demonstrated the major 
improvement of the ratio is caused by the use of realistic dipole cross 
section.  
We have also discussed $t$-slope of the diffractive leptoproduction 
of $J / \psi$ and $\psi '$.  
We have emphasized that the difference of $B_{J /\psi} - B_{\psi '}$ 
is sensitive to the shape of the dipole cross section.  
Our result for the photoproduction is about 
$0.3 \mbox{GeV}^{-2}$, which 
seems to be consistent to recent preliminary data.

\end{document}